\documentclass[a4paper, BackFigs,11pt]{article}
\usepackage{amsfonts}
\usepackage{amsmath}
\usepackage{amssymb}
\usepackage{graphicx}%
\usepackage{color}
\usepackage{float}
\usepackage[numbers,sort&compress]{natbib}
\begin{document}

\title{Broadband perfect light trapping in the thinnest monolayer graphene-MoS$_{2}$ photovoltaic cell}

\author{\small Yun-Beng Wu$^{1}$, Weng Yang$^{2}$, Tong-Biao Wang$^{3}$, Xin-Hua Deng$^{3}$, Jiang-Tao Liu$^{1,3*}$\\  \footnotesize
$^{1}$Nanoscale Science and Technology  Laboratory,  Institute for Advanced Study, \\
 \footnotesize  Nanchang University, Nanchang 330031, China \\  \footnotesize  $^{2}$Beijing Computational Science Research Center, Beijing 100084, China\\
 % For a paper whose authors are all at the same institution,
 % omit the following lines up until the closing ``}''.
 % Additional authors and addresses can be added with ``\and'',
 % just like the second author.
 \footnotesize $^{3}$
Department of Physics, Nanchang University,
Nanchang
330031, China\\
 \footnotesize   $^{*}$Email: jtliu@semi.ac.cn
 }

\maketitle
\begin{abstract}
The light absorption of a monolayer graphene-molybdenum disulfide photovoltaic (GM-PV) cell in a wedge-shaped microcavity with a spectrum-splitting structure is investigated theoretically. The GM-PV cell, which is three times thinner than the traditional photovoltaic cell, exhibits up to 98\% light absorptivity in a wide wavelength range. This rate exceeds the fundamental limit of nanophotonic light trapping in solar cells. The effects of defect layer thickness, GM-PV cell position in the microcavity, incident angle, and lens aberration on the light absorption rate of the GM-PV cell is explored. Regardless of errors, the GM-PV cell can still achieve at least 90\% light absorptivity with the current technology. Our proposal provides different methods to design light-trapping structures and apply spectrum-splitting systems.
\end{abstract}

\section{Introduction }
Nano-sized photovoltaic cells have gained significant attention because of their low cost and high efficient\cite{B13IMD, JAP12SM,PNAS10ZFY,LSA14CFG, S13JW, NC13GM, NP13PK, NL10AJN, SEMSC14SS, NL13MB,NL14MG,AEM14CB,N14ZYH,SR14LZ}. Working medium materials such as monocrystalline Si account for over 40\% of the total cost of a traditional photovoltaic cell. Therefore, reducing the thickness of the working medium decreases the cost of photovoltaic cells. In optoelectronic devices such as photovoltaic cells and photoelectric detectors, the thickness of the working medium layer should be lower than the diffusion length of the carriers to reduce the losses induced by photon-generated carrier recombination\cite{ B13IMD,JAP12SM}. Nano-sized photovoltaic cells can be synthesized using materials with low mobility, such as amorphous silicon and organic semiconductor materials, to further reduce the cost\cite{B13IMD,JAP12SM}. Decreasing the thickness of photovoltaic cell can also reduce the losses induced by photon-generated carrier recombination and thus enhance photoelectric conversion efficiency. Bernardi et al. proposed an ultrathin photovoltaic cell (i.e., about 1 nm thick) composed of monolayer graphene and monolayer MoS$_{2}$ \cite{NL13MB}. The light conversion efficiency per unit of mass of this photovoltaic cell is three times higher that of the traditional photovoltaic cell. Since the discovery of this ultrathin photovoltaic cell, different types of photovoltaic cells based on two-dimensional (2D) materials have been proposed \cite{NL14DJG, NL14MMF, N14MS, AN14MLT, JPCL14JD}. However, the light absorptivity of the monolayer graphene-molybdenum photovoltaic cell (GM-PV cell) is only about 10\%. Thus, A light-trapping structure should be employed to optimize the light absorption of GM-PV cell.

 In general, a light-trapping structure is not designed on the basis of light interference and has a poor light-trapping efficiency. Broadband perfect absorption can hardly be achieved in a $<$300 nm-thick medium layer because of the limitations of Lambertian light trapping \cite{B13IMD,JAP12SM,PNAS10ZFY,LSA14CFG}. Moreover, an anti-reflection layer with the thickness of several micrometers should be added \cite{B13IMD,JAP12SM,PNAS10ZFY,LSA14CFG}. A strong light localization can be realized based on an interference-based light-trapping structure (i-LTS), which can greatly enhance the light absorption and simultaneously reduce the reflection of a semiconductor\cite{PNAS10ZFY,LSA14CFG,EES14KXW,NM12KV,NC13ERM,OE12MYK,AEM14SL,JAP13MR,AM13QG,N14CHC,SR14LBL,OE10SBM}. Therefore, an additional anti-reflection layer is no longer needed. However, limited to the fundamental limit of nanophotonic light trapping in solar cells\cite{JAP12SM,PNAS10ZFY},   a strong absorption can hardly be achieved at the full spectrum of the sunlight. However, the resonant frequency of i-LTS can usually be adjusted by changing a characteristic parameter, such as the cavity length and the lattice constant of the photonic crystal. Therefore, employing a spectrum-splitting structure can focus sunlight with different wavelengths onto a light-trapping structure with different characteristic parameters to enhance light absorption through resonance in a wide frequency range.

Spectrum-splitting structure is widely applied in super-efficient photovoltaic cells\cite{B98AG,NM12AP,SEM04AGI,PP09AB,PP10MAG}. This structure can reduce the thermal losses caused by the mismatch between the photon energy and the semiconductor's band gap by focusing light with different wavelengths on semiconductors with various band gaps. Thermal losses refer to the thermal energy that is converted from the part of the photon energy that is greater than the band gap and photons with a lesser energy than the semiconductor's band gap cannot be absorbed by the semiconductor \cite{B98AG,NM12AP,SEM04AGI,PP09AB,PP10MAG}. Thermal losses may be reduced to approximately 10\% by using 8-10 semiconductors with different band gaps to constitute a photovoltaic cell \cite{NM12AP}.
In our previous studies, we combined the spectrum-splitting structure and the resonance back-reflection light trapping structure to achieve broadband perfect absorption in a semiconductor film with a thickness of approximately 100 nm\cite{ PCCP15JTL}. However, light localization is weak in the resonance back-reflection light trapping structure. Hence, perfect light absorption can hardly be achieved in a 1 nm-thick medium layer. Demands on material mobility can be minimized by reducing the thickness of the medium layer to $\sim$1 nm, and the energy band structure and optical properties of a $\sim$1 nm semiconductor film can be controlled by adjusting structural and physical parameters such as stress. This process promotes the applications of energy band engineering in photovoltaic cells.

Recent many studies have focused on exploring the enhanced absorption of light in 2D materials such as graphene by using various optical microstructures \cite{PRL12ST, PRB12AF, OL13MAV,NL12MF, NC12ME,AP14JZ,PRB08ZZZ, APL13XZ, APL13QY, AP15RY, NL15DHL}, e.g., perfect light absorption can be achieved in monolayer graphene by using a asymmetry of photonic crystal microcavity within a narrow frequency range. In the present study, we combined photonic crystal microcavity with a wedge-shaped defect layer and spectrum-splitting structure. Using this structure produced a photovoltaic cell that is three times thinner than the traditional photovoltaic cell and that exhibits a preferable light absorptivity higher than 98\% in a wide wavelength range. We also determined the effects of defect layer thickness, GM-PV cell position in the microcavity, incident angle, and lens aberration on the light absorption rate of the GM-PV cell to compare the calculation results and the actual outcome. Regardless of errors, the light absorptivity of the GM-PV cell can exceed 90\% with the current technology. The study provides different methods to design light-trapping structures and apply spectrum-splitting systems.

\section{NUMERICAL RESULTS}

\begin{figure}[H]
\includegraphics[width=7cm,clip]{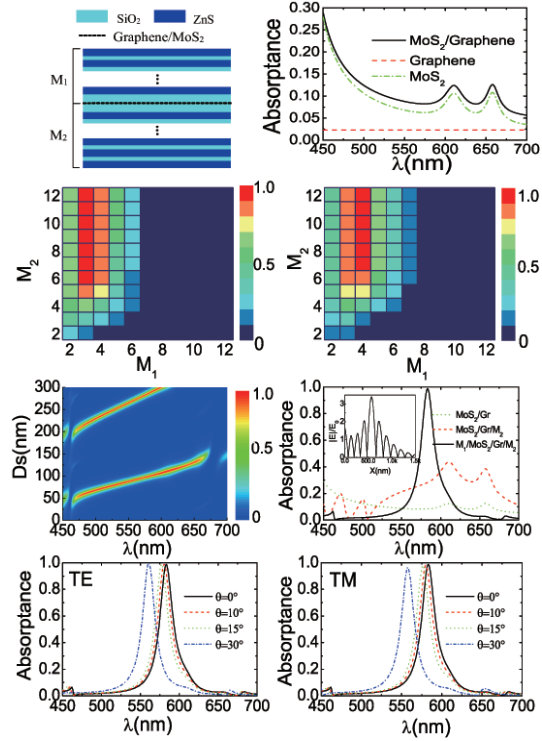}\centering
\caption{(Color online)(a) Schematic of a common photonic crystal microcavity. (b) Light absorptance of the monolayer graphene, MoS$_{2}$, and GM-PV cell. (c) Changes in absorptance with the periodicity of the distributed Bragg reflectors on both sides of the microcavity at wavelengths of (c) 470 nm and (d) 610 nm. (e) Absorptance variation of GM-PV cell in the microcavity at different wavelengths (inset: light field distribution). (f) Contour chart of the absorptance of GM-PV cell at different wavelengths and of the thickness of the defect layer. Absorptance variation of the GM-PV cell with changing wavelengths at different incident angles: (g) TE mode and (h) TM mode.
}
\label{fig1}%
\end{figure}

The effect of the common microcavity on the absorptance of the GM-PV cell was studied firstly. Fig. 1(a) shows the structure of a common microcavity; the defect layer with  dielectric $n_{c}$ is in the middle of the cavity, and the GM-PV cell is at the center of the defect layer. The distributed Bragg reflectors (DBRs) in which two different   material with different dielectric ($n_{1}$ and $n_{2}$)
are alternately distributed are on both sides of the cavity with periodicities of  $M_{1}$ and  $M_{2}$. $n_{c}=n_{1}=1.55$  (e.g., SiO$_{2}$) and $n_{1}=2.59$ (e.g., ZnS) are used in the calculation. All layers are nonmagnetic ($\mu$ =1). For a direct comparison, the light absorptivities of the monolayer graphene, MoS$_{2}$, and GM-PV cell were calculated. The refraction
index of graphene $n_{g}=3.0+\mathcal{C}_{1}\frac{\lambda}{3}i$, where $\mathcal{C}_{1}=5.446$ $\mu m ^{-1}$, the thickness of the graphene monolayer $d_{g}=0.34nm$ \cite{APL09MB}. The data of MoS$_{2}$ are taken from Ref. \cite{JAP14JTL}. The results are presented in Fig. 1(b). The absorptance of the GM-PV cell is approximately 10\%, increases rapidly at short wavelengths, and reaches 20\%. Previous studies reported that using an asymmetric microcavity can reduce reflection and enhance light absorption\cite{OL13MAV}. We calculated the changes in absorptance at wavelengths of 470 and 610 nm [Figs. 1(c) and 1(d)] as the periodicity of the DBRs on both sides of the microcavity varies. At 470 nm, the GM-PV cell exhibits a high absorptance, and the required Q-value is small. When $M_{1}=2$ and $M_{2}>5$, the absorption is almost saturated. At 610 nm, the GM-PV cell exhibits a low absorptance, and the required Q-value is large. When $M_{1}=3$ and $M_{2}>6$, the absorption is basically saturated. The changes in absorptance with varying wavelengths are presented in Fig. 1(e). The maximum absorptance is 98.5\%, and the full width at half maximum (FWHM) is 20.2 nm. In this structure, the absorptance is close to 1 is because interference can reduce the reflection, and light can be reflected in the microcavity for multiple times causing a strong light localization [inset of Fig. 1(e)].

The resonance state in the microcavity is sensitive to the thickness of the defect layer and the incident angle. The calculated results are illustrated in Figs. 1(f)-1(h). The resonant wavelength of the microcavity satisfies $m_{i}\lambda_{c}/2=L_{c}\cos\theta'$, where $L_{c}=n_{c}d_{c}$  denotes the optical path of the microcavity, $n_{c}$ and $d_{c}$ is the refractive index and the thickness of the defect layer, $m_{i}$ is the positive integer, and $\theta'=\arcsin\theta_{i}/n_{c}$  signifies the propagation angle of the light in the defect layer, $\theta_{i}$ is the incident angle. Thus, the resonant wavelength linearly increases as the defect layer increases in thickness. Then, the resonant wavelength moves toward the direction of the short wave as the incident angle increases. Perfect absorption is mainly distributed at the photonic band gap (470-670 nm) because the thickness of the photonic crystals on both sides remains unchanged. The resonant wavelength also correlates with the propagation angle in the defect layer. The higher the refractive index of the defect layer, the smaller the propagation angle. Accordingly, minimal changes in the resonance peak occur as the incident angle varies. In the calculations, the resonance peak is sensitive to the incident angle when the refractive index of the defect layer is 1.55. When the incident angle is $15^{\circ}$, the resonance absorption peak slightly shifts. Comparatively, the resonance peak shifts with a distance up to a FWHM when the incident angle is $30^{\circ}$.

Basing from the calculation results, we can conclude that the light absorption in a common photonic crystal microcavity can only be enhanced at some specific wavelengths. In other words, the method cannot be directly applied in photovoltaic cells. However, the microcavity-based resonant wavelength linearly increases as the thickness of the defect layer increases. Therefore, a wedge-shaped defect layer is introduced, and sunlight with different wavelengths can be focused onto different positions in the microcavity through the spectrum-splitting structure. The resonance absorption of light can be achieved at the position where light is focused because of the appropriate thickness distribution of the defect layer. Accordingly, resonance-induced absorption enhancement can be realized in a broad frequency range [Fig. 2(a)]. Similarly, the DBRs on both sides of the microcavity are also wedge-shaped to further increase the band gap [Fig. 2(b)].

\begin{figure}[H]
\includegraphics[width=7cm,clip]{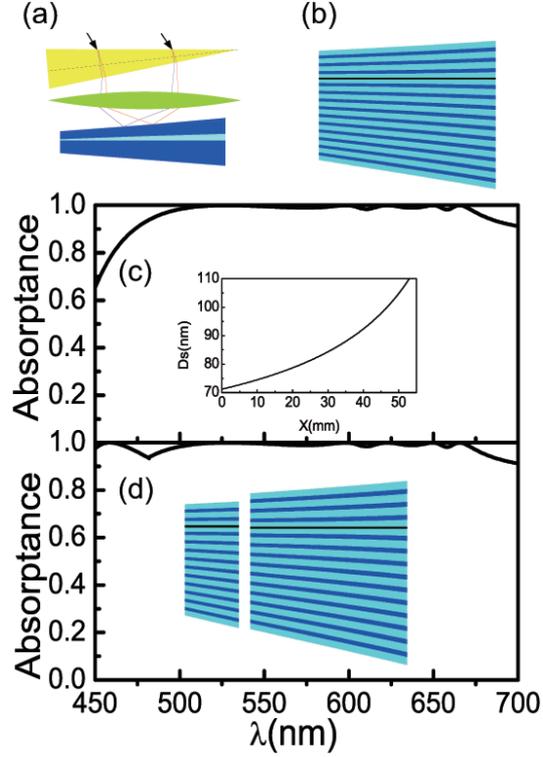}\centering
\caption{(Color online)Schematic of (a) Spectrum-splitting system and (b) Wedge-shaped photonic crystal; (c) absorptance of the GM-PV cell in the wedge-shaped microcavity (inset: thickness variation of the defect layer with varying position coordinates); (d) absorptance of the GM-PV cell in the split wedge-shaped microcavity (inset: shoulder-to-shoulder wedge-shaped microcavity).
}
\label{fig2}%
\end{figure}

Fig. 2(c) presents the calculation results, in which the periodicities of the upper and lower DBRs are 3 and 10, respectively. Since the slope of the  wedge-shaped layer is smaller than $1.7\times10^{-6}$, the layers can be
treated as parallel planes at each point. Thus, the transfer matrix
method can be used in the calculation. Light with different wavelength incidents on different positions in the microcavity through the spectrum-splitting structure can achieve a resonance-induced absorption enhancement by adjusting the thickness of the defect layer. The calculated thickness distribution of the resonance defect layer is presented in the inset of Fig. 2(c). In this system, the absorptance of the GM-PV cell can exceed 98\% at a wavelength of 500-670 nm (i.e., 670 nm at approximately the band gap of the monolayer MoS$_{2}$) by combining the spectrum-splitting structure and i-LTS. The absorptance of the GM-PV cell significantly decreases when the wavelength is less than 500 nm. This result can be attributed to the strong absorption of the GM-PV cell at short wavelengths and the large periodicity of the DBR. These features can suppress interference, increase reflection, and reduce absorptance [Fig. 1(c)]. Nevertheless, shoulder-to-shoulder microcavities with different Q-values can be used in the design [inset of Fig. 2(d)] similar to the design of the shoulder-to-shoulder sub-batteries adopted in common spectrum-splitting systems. Light absorption at wavelengths less than 500 nm can be enhanced by reducing the periodicity of the upper DBR [Fig. 2(d)]. By using the shoulder-to-shoulder microcavities, the DBR can also fabricated with parallel layers \cite{ppp}.

To associate the calculation results with practical production, the effects of GM-PV cell location in the microcavity, defect layer thickness, incident angle, and lens aberration on the absorptance of the GM-PV cell were further investigated. The calculation results are presented in Fig. 3. Some other effects are shown in the supplementary materials \cite{ppp}

\begin{figure}[H]
\includegraphics[width=7cm,clip]{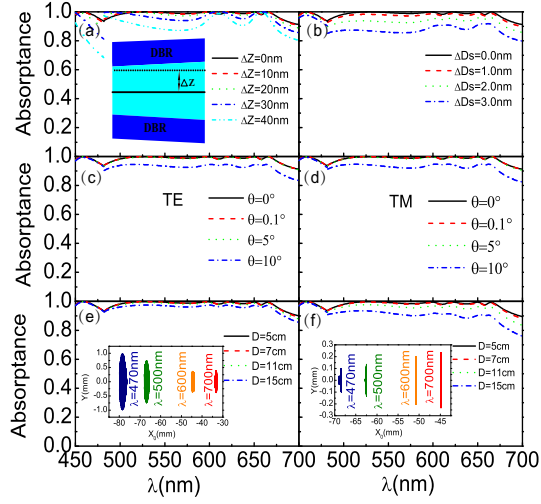}\centering
\caption{(Color online)Absorptance of GM-PV cell at (a) different off-center positions of the GM-PV cell in the microcavity; (b) different defect layer thicknesses; (c) different angles (corresponding to TE mode); (d) different angles (corresponding to TM mode); (e) different lens diameters when the vertex angle of the prism is  $\alpha_{A}=45\circ$; (f) different lens diameters when the vertex angle of the prism is  $\alpha_{A}=30\circ$; (g) different photonic crystal layer thicknesses with a constant period and optical path; (h) randomly changed photonic crystal layer thickness.
}
\label{fig3}%
\end{figure}

Light localization is the strongest at the center of the cavity. The absorptance of the GM-PV cell declines when it is off-center. Nevertheless, the absorptance at the off-center position   is slightly affected because the light field is almost evenly distributed in the cavity. At  $\Delta x=10$ nm, the absorptance of the GM-PV cell exhibits almost no variation. At   $\Delta x=20$ nm, the absorptance slightly varies. At  $\Delta x\geq30$ nm, the absorptance significantly declines. The absorptance change increases at short wavelengths. This result can be attributed to the fact that a thinner defect layer at short wavelengths leads to a greater relative deviation. A change in defect layer thickness influences the absorptance of the GM-PV cell. The resonant frequency of the microcavity varies as the thickness of the defect layer varies. Thus, the resonance-induced perfect absorption of the light incident on the cavity can no longer be achieved, i.e., the absorptance is reduced. The absorptance slightly changes when the variation amplitude of the defect layer thickness is 1 nm. The light absorptance at $>500$ nm wavelengths significantly decreases when the variation amplitude of the thickness is 2 nm. However, it can still exceed 90\%. The light absorptance at $<500$ nm wavelengths shows almost no change due to the lower Q-value of the microcavity. The light localization becomes weaker and the FWHM increases as the Q-value  decreases.    The tolerance of film growth can be less than 1 nm or even as low as the thickness of an atomic layer because of the development of material growth techniques such as molecular beam epitaxy (MBE).

As described earlier, varying incident angles can also change the resonant frequency of the microcavity. Nevertheless, the resonant frequency of the microcavity slightly varies when the incident angle slightly changes. The calculated results are presented in Figs. 3(c) and 3(d). The absorptance slightly changes when the variation amplitude of the incident angle is $5^{\circ}$ but significantly declines when the incident angle is altered by $10^{\circ}$. The incident angle can be controlled at approximately $0.1^{\circ}$ by using the ray tracing system\cite{RE14YY}.

Aberration causes the appearance of a light spot of a certain size after lens imaging. The wavelength perfectly matches with the thickness of the defect layer around the center of the light spot, whereas the wavelength in other positions cannot be matched properly, leading to a decline in absorptance. Figs. (e) and (f) present the effects of aberration on the absorptance of the GM-PV cell. Larger lens diameter means larger light spot. Accordingly, mismatch becomes evident, and GM-PV cell absorptance is reduced. The effect of aberration on the absorptance of the GM-PV cell mainly depends on the Q-value and the light-splitting ability of the spectrum-splitting structure. Smaller Q-value means weaker light localization and larger FWHM. Accordingly, the effect of aberration is weak under this condition. In lenses with varying diameters, the absorptance variation at wavelengths less than 500 nm significantly differs from that at wavelengths higher than 500 nm. Fig. (e) and Fig. (f) show that larger vertical angle of the prism means stronger light-splitting ability and smaller variation gradient. That is, the effect of aberration is minimal.

\section{Discussion and conclusions}

\textbf{Spectrum-splitting system.} The spectrum-splitting structure can reduce the thermal losses caused by the mismatch between the photon energy and the semiconductor's band gap. It can also enhance the light-trapping efficiency via combined utilization with i-LTS. The introduction of the spectrum-splitting structure increases the flexibility of the design of the light-trapping structure. The design of the shoulder-to-shoulder battery can also be adopted. Various light-trapping structures can be selected for sub-batteries with different wavelengths and materials. In addition to the microcavity in our current study, other light-trapping structures can also be employed and combined with the spectrum-splitting structure. These structures include photonic crystal, plasma, and nano-semiconductor-column array with resonant frequencies that are continuously adjustable by some structural parameters. The spectrum-splitting structure studied in this article is composed of disperse prism and convex lens. The structure is simple and can be manufactured easily, which is beneficial to revealing the fundamental principles. However, the light-splitting efficiency is not high and the system can reflect 15\%-20\% of sunlight. In this case, highly efficient spectrum-splitting structure should be adopted\cite{NM12AP,SEM04AGI,PP09AB,PP10MAG}. The disperse plane lens may be a potential choice \cite{NL12FA}.

\textbf{Light-trapping ability and technique precision.} Theoretically, it is no limitation in i-LTS combined with the spectrum-splitting structure. But the practical light-trapping ability is limited to the technique precision.  The light-trapping ability of an interference structure can be measured by the Q-value. Larger Q-value means stronger light-trapping ability. Although the Q-value of optical microstructures without absorption can reach up to $10^{8}$, the light-trapping ability is limited by the processing technique. Basing from the definition of Q-value  $Q=\omega_{0}/\Gamma$, we can conclude that greater Q-values mean smaller FWHM. Meanwhile, smaller FWHM indicates higher demands on the processing precision. For example, with the assumption that the resonant wavelength of the microcavity is  $\lambda_{c}=2L_{c}=2n_{c}d_{c}$, resonant wavelength will be changed by $\Delta\lambda_{c}=2n_{c}\Delta d_{c}$  if  $d_{c}$ is varied to $d_{c}+\Delta d_{c}$  because of processing errors, and the absorptance will be reduced by 50\% if  $\Delta\lambda_{c}$ reaches half of the FMHM. Thus, after considering processing precision, the light-trapping ability based on the microcavity can be increased by at most two to three times than the results of the study.

\textbf{Transparent electrode.} On the basis of the strong light localization in i-LTS, the light absorptance of the transparent electrode can also be enhanced. Thus, the extinction coefficient of the transparent electrode should be lower. The primary problem with the transparent electrode in the traditional photovoltaic cell is that it is required to maintain a small degree of light absorption within the entire spectral region of sunlight. However, different transparent electrodes can be used on sub-batteries by adopting the shoulder-to-shoulder design of the sub-batteries after using spectrum-splitting structure. These transparent electrodes are only required to exhibit minimal light absorption within a specific frequency range, i.e., the difficulty in the design of the transparent electrode can be significantly reduced. That is also why it is difficult to obtain the perfect absorption in monolayer graphene with microcavity\cite{NC12ME,NL12MF}.

\textbf{Materials.} The thinnest photovoltaic cell (i.e., the graphene-MoS$_{2}$ photovoltaic cell) was employed in this work to investigate the light-trapping ability of the structure combined with spectrum-splitting structure and i-LTS. The requirements on material mobility can be minimized because the thickness of the working medium layer is only 1 nm. This phenomenon promotes the application of materials with low mobility. Such materials include noncrystalline materials and organic materials in photovoltaic cells. When the quantum tunneling effect is considered, the insulating medium with a low barrier height can be used to prepare the absorption medium. In particular, when the thickness of the working medium is reduced to a few nanometer, the energy band structure and optical properties can be adjusted by changing physical and structural parameters, such as stress. This step can largely promote the application of energy-band engineering in photovoltaic cells.

\textbf{Feasibility of the experiments.}  The structures can achieve a high light absorptance after considering the present technique precision. The combined utilization of microcavity and the traditional semiconductor microstructures, such as quantum well, is quite mature. However, the combined utilization of microcavity and 2D materials remains a challenge, but a photonic crystal microcavity that contains graphene has been produced. The primary difficulty in fabricating this product is integrating i-LTS with spectrum-splitting structure to establish the specific wedge-shaped medium layer. Prineas et al. promoted the growth of 210 wedge-shaped semiconductor layers through MBE to reduce the speed of light\cite{APL06JPP}.

\textbf{Limitations.} Similar to the traditional spectrum-splitting structure-based photovoltaic cells, this type of photovoltaic cell is applicable only under direct sunlight.  Under indirect sunlight, the conversion efficiency is significantly reduced and the utilization ratio of the scattered sunlight is low. Solar tracker and feedback mechanical system should be used.  In regions with a short period of direct sunlight exposure, the GM-PV cell should be used simultaneously with traditional photovoltaic cells.

In conclusion, broadband perfect absorption can be achieved in the 1 nm-thick GM-PV cell by combining spectrum-splitting structure and i-LTS. The achieved light absorptance exceeds the fundamental limit of nanophotonic light trapping in solar cells. Regardless of errors in defect layer thickness, GM-PV cell position in the microcavity, incident angle, and lens aberration, the light absorptance of the GM-PV cell manufactured with the current technology can still exceed 90\%. This study not only provides new different way to design light-trapping structures and apply spectrum-splitting systems but also presents significant application prospects in the development of ultrathin and highly efficient photovoltaic cells.

\section*{Acknowledgements}
We would like to thank Nianhua Liu for fruitful discussion.
This work was supported by the NSFC (Grant No. 11364033,
11274036, 11322542, and 11264029), the MOST (Grant No.
2014CB848700), and Science and Technology Project of the
Education Department of Jiangxi Province (Grant No. GJJ13005).

\newpage

%\centering{ \section{Supplementary Materials}}
%The study demonstrated that the THz absorption by monolayer MoS$_{2}$ is very small over a broadband frequency range even under high carrier concentration and large incident angle. Its equivalent loss is one to three grades smaller than that of graphene. The transmission of monolayer MoS$_{2}$  is much larger than that of traditional GaAs or InAs electron gas. We studied the field-effect tubular structure of monolayer MoS$_{2}$-insulation layer-graphene, which allows the THz absorption of graphene to reach saturation even under low voltage even while the THz absorption of monolayer MoS$_{2}$ is only approximately 5\% at the maximum. Therefore, monolayer MoS$_{2}$ can be used to make transparent electrodes within the THz frequency range. This development has important prospective applications in optoelectronic devices within the THz frequency range.

\end{document}